\documentstyle[12pt]{article}
\newcommand{\alphaa}{\mbox{\boldmath $\alpha$}}
\newcommand{\betaa}{\mbox{\boldmath $\beta$}}

\begin{document}
\baselineskip=22pt plus 0.2pt minus 0.2pt
\lineskip=22pt plus 0.2pt minus 0.2pt
\font\bigbf=cmbx10 scaled\magstep3
\begin{center}
 {\bigbf Gravitons from a loop representation of linearised gravity}\\

\vspace*{0.35in}

\large

Madhavan Varadarajan
\vspace*{0.25in}

\normalsize

{\sl Raman Research Institute,
Bangalore 560 080, India and Physik Department E22,
Technische Universit{\" a}t M{\" u}nchen, 85748 Garching,
Germany.}
\\
madhavan@rri.res.in\\
\vspace{.5in}
April 2002\\
\vspace{.5in}
ABSTRACT

Loop quantum gravity is based on a classical formulation of 3+1 gravity in 
terms of a real $SU(2)$ connection. Linearization of this classical 
formulation about a flat background 
 yields a description of linearised gravity in terms of a {\em real} 
$U(1)\times U(1)\times U(1)$
connection. A `loop' representation, in which holonomies of this connection
are unitary operators, can be constructed.
These holonomies are not well defined operators 
in the standard graviton Fock representation. We generalise our recent
work on photons and $U(1)$ holonomies to show that Fock space gravitons 
are associated with distributional states in the $U(1)\times U(1)\times U(1)$ 
loop representation. Our results may  illuminate certain aspects of the 
much deeper (and as yet unkown,) relation between 
 gravitons  and states in
nonperturbative loop quantum gravity.

This work  leans heavily on earlier seminal  work by 
Ashtekar, Rovelli and Smolin (ARS)
on the loop representation of linearised gravity using {\em complex}
connections. In the last part of this work, we show that the 
loop representation based on the {\em real} $U(1)\times U(1)\times U(1)$ 
connection also 
provides a useful kinematic arena in which it is possible to
express the ARS complex connection- based results in the mathematically
precise language currently used in the field.

\end{center}

\pagebreak

\setcounter{page}{1}

\section*{1. Introduction} 

Loop quantum gravity \cite{aareview,ttreview,carloreview} 
is an attempt to apply standard Dirac quantization 
techniques to a classical Hamiltonian formulation 
of  3+1 gravity in which the 
basic variables
are a spatial $SU(2)$ connection and its conjugate triad field. 
In addition to the usual diffeomorphism and Hamiltonian constraints, 
the formulation also has an $SU(2)$ Gauss law constraint which ensures that 
triad rotations are gauge. At the  $SU(2)$ gauge invariant level (also 
referred to as  the kinematic level), the representation 
space is generated by the action of (traces of) holonomies of the 
connection on a cyclic state. Since holonomies are labelled by 1 dimensional,
arbitrarily complicated  loops, the basic quantum excitations may be 
visualised as 1 dimensional and `polymer- like'. Physical states, 
which are in the  kernel of all the constraints, are expressible as 
certain kinematically non- normalizable, linear combinations of these
polymer-like excitations \cite{ALM^2T}. 

A key open question is:
how do classical configurations of the gravitational field arise ? 
In particular, how does flat spacetime (and small 
perturbations around it) arise from non-perturbative quantum states of the  
gravitational field?
The latter question is particularly interesting for the following reason.
Small perturbations about flat spacetime correspond to solutions of 
linearized gravity. Quantum states of linearised gravity lie in the 
familiar graviton Fock space  on which the conventional perturbative 
approaches to quantum gravity are based. Such approaches seem to fail due to
nonrenormalizability  problems. Thus, an understanding of the relation between
the quantum states of linearised gravity and states in full nonperturbative
loop quantum gravity would shed light on the reasons behind the failure of
perturbative methods. 

In this work we focus exclusively on understanding certain structures in 
quantum linearised gravity, which concievably (but by no means, assuredly!) 
could play a role in the much deeper issue of the relation between 
perturbative and non-perturbative states.
Our starting point is the linearization of the classical $SU(2)$ formulation 
\cite{barbero} 
on which loop quantum gravity is based. 
This linearization is described in section 2 wherein we also show that 
 the linearized
Gauss Law constraint generates $U(1)\times U(1)\times U(1)$ transformations
on the linearized connection.

Using the methods of 
\cite{AI,AL,Corichi-Krasnov,AI2,medual},  $U(1)^3$ counterparts of the 
$SU(2)$ based structures of loop quantum gravity can be constructed.
In particular, at the  $U(1)^3$ gauge invariant level, a `kinematic' Hilbert
space  ${\cal H}_{kin}$ exists which is spanned by 1 dimensional polymer-like 
excitations associated with (triplets \cite{lingrav} of) loops. Holonomies
of the linearised connection are represented as unitary operators on 
${\cal H}_{kin}$. We exhibit this representation in section 3a.

As realised in \cite{lingrav}, the operator corresponding to the 
magnetic field of the linearised connection plays a key role in 
expressing the
linearised diffeomorphism and Hamiltonian constraints as quantum operators.
It turns out that this operator is not well defined in 
${\cal H}_{kin}$. Neverthless it can be represented on a vector space 
$\Phi^{*L}_{kin}$ of appropriately well behaved distributional combinations
of elements in ${\cal H}_{kin}$. Using this representation of the 
magnetic field operator, we identify 
 the kernel, $\Phi^{*L}_{phys}$, 
 of all the constraints. 
Since  $\Phi^{*L}_{phys}\subset \Phi^{*L}_{kin}$,
elements of $\Phi^{*L}_{phys}$  are also associated with infinite, 
kinematically non-normalizable sums of 1 dimensional polymer-like excitations.
Section 3b is devoted to a discussion of the magnetic field operator and 
an evaluation of the kernel of the quantum constraints.

The standard graviton Fock space representation of linearised gravity is very 
different from the above `loop' representation. The basic excitations in 
Fock space are  3d and wavelike in contrast to the polymer-like nature of 
excitations in the loop representation. Moreover, in the Fock representation 
the connection is an operator valued distribution which needs to be smeared 
in 3 dimensions to obtain a well defined operator. Since holonomies involve 
only 1 dimensional smearings along loops, they are not well defined operators
on Fock space.

In view of the above remarks, it is a non-trivial task to relate Fock space
gravitons to elements in $\Phi^{*L}_{phys}$.
In section 4, we generalise the considerations of \cite{medual} to  
relate the loop representation of linearised gravity to its standard Fock 
representation. As in \cite{medual} we use the Poincare invariance of the 
Fock vacuum to identify graviton states in $\Phi^{*L}_{phys}$.
\footnote{In this paragraph (but {\em not} in section 4) 
we gloss over the very important difference between the standard
Fock representation and the $r$- Fock representation \cite{medual} 
for reasons of brevity and pedagogy.}

%In section 4 we define a 
%`Fock' inner product on the subset of 
%$\Phi^{*(L)}_{phys}$ corresponding to gravitons
% and show that the reality conditions
%on Dirac observables are implemented as adjointness conditions on 
%the associated operators under this inner product.

Recall that the starting point of this work is the linearization of a
{\em real} $SU(2)$ formulation \cite{barbero,QSD} of classical gravity.
 The basic variable is a real $SU(2)$ connection and the associated 
Barbero-Immirzi parameter \cite{barbero,immirzi} is real. In contrast
Ashtekar, Rovelli and Smolin use the complex self dual Ashtekar-Sen connection 
\cite{selfdual} in their pioneering work \cite{lingrav} on a 
loop representation
of linearised gravity. 
This corresponds to the choice of  an imaginary Barbero-Immirzi 
parameter. In section 5, 
we show how to extend the  considerations of sections 2- 4 
to the case of an arbitrary  complex 
Barbero- Immirzi parameter. Section 6 is devoted to a 
discussion of our results.

As mentioned above, our real interest is in the deeper issue of the relation
between states in linearised gravity and in full quantum gravity rather than
just in structures in linearised gravity. One possible way to approach the 
deeper issue is to divide it into two parts. First, 
since both loop 
quantum gravity and the $U(1)^3$ representation are 
structurally similar, we may try to relate the two. 
This is the really hard part.
The second 
(and much easier) part is to relate the $U(1)^3$ representation 
to the standard graviton Fock representation. It is only the second
part that we accomplish in this paper.

This work  is heavily based on  the Ashtekar-Rovelli-Smolin paper 
\cite{lingrav}
 and on \cite{medual}. For this reason, we shall be
very brief in our presentation and sketch only the important points. The 
reader may consult \cite{lingrav,medual} for more details.
Indeed, this work  may be read as a mathematically precise formulation 
of the earlier ARS \cite{lingrav}
 work in the context of the  subsequent developments in the 
field as reflected in, for example, 
\cite{ALM^2T,QSD,AL,AI,M^2,Baezproj,ALproj,RSspinnet,
Baezspinnet,medual,mefock}.

We use units in which Newton's constant, Planck's constant and the speed of 
light are unity.

\section*{2. Classical linearised gravity 
as theory of $U(1)\times U(1) \times U(1)$
connections.}

Our starting point is 
the  Hamiltonian formulation of 3+1 gravity discussed in \cite{barbero}
The spacetime 
manifold has topology $\Sigma \times R$ where  $\Sigma$  is a
3 dimensional orientable manifold. The phase space variables are a spatial 
$SU(2)$ connection, $A_a^i ({\vec x})$ and a densitized triad field 
$E^b_j({\vec y})$. 
Here
$a,b$ denote spatial components, $i,j$ denote internal $SU(2)$ Lie algebra
components and ${\vec x},{\vec y}$ denote (in general, local) coordinates
on $\Sigma$. The only non- vanishing Poisson bracket is 
\begin{equation}
\{ A_a^i ({\vec x}), E^b_j({\vec y}) \} = 
    {\gamma_0 \over 2}\delta^b_a \delta^i_j 
               \delta ({\vec x}, {\vec y}) .
\label{pb}
\end{equation}

Here, $\gamma_0$ is the Barbero-Immirzi parameter \cite{barbero,immirzi}.
The spin connection associated with the triad field is denoted by $\Gamma^i_a$,
the curvature of $A_a^i$ by $F_{ab}^i$ and the gauge covariant derivative 
associated with $A_a^i$ by ${\cal D}_a$.  The 
constraints of the theory are the Gauss law constraint $G_i$, the 
vector or diffeomorphism constraint $V_a$ and the Hamiltonian constraint
$C$. They are given by
\begin{eqnarray}
G_i &=& {\cal D}_a E^a_i , 
\label{gauss}\\
V_a &=& E^a_iF_{ab}^i,
\label{vector} \\
C &=& \epsilon^{ijk} E^a_iE^b_j F_{abk}- 2{\gamma_0^2 +1\over \gamma_0^2}
 E^a_{[i}E^b_{j]}(A_a^i-\Gamma_a^i)(A_a^j- \Gamma_a^j).    
\label{scalar}
\end{eqnarray}
 
The $SU(2)$ variables are related to the ADM variables as follows. The 
densitized triad and the 3- metric, $q_{ab}$, are related through 
\begin{equation}
q q^{ab} = E^{ai}E^b_i
\label{qab}
\end{equation}
where $q$ is the determinant of  $q_{ab}$. When $G_i =0$ the extrinsic
curvature, $K_{ab}$,  can be extracted from the $SU(2)$ variables through
\begin{equation}
\gamma_0 K_{ab}E^b_i = {\sqrt {q}}(A_{a}^i -\Gamma_a^i) .
\label{kab}
\end{equation}

To define the linearised theory about a flat background we choose $\Sigma=R^3$
and fix, once and for all, a cartesian coordinate system $\{ {\vec x}\}$
as well as an orthonormal basis in the Lie algebra of $SU(2)$. Henceforth
all components  refer to  this cartesian coordinate system 
and to this internal basis. We linearise the $SU(2)$ formulation about the 
phase space point $(A_a^i=0, E^a_i = \delta^a_i)$. As in \cite{lingrav},
we denote the fluctuation in the triad field by $e^a_i$  so that 
\begin{equation}
E^a_i = \delta^a_i + e^a_i .
\label{ltriad}
\end{equation}
Since the background connection vanishes, there is no need to introduce a new
symbol for the fluctuation in the connection. The Poisson brackets between the
linearised variables are  induced from (\ref{pb}). The only non-vanishing
Poisson bracket is
\begin{equation}
\{A_a^i({\vec x}), e^b_j({\vec y})\}_L = {\gamma_0 \over 2} 
         \delta^b_a \delta^i_j 
               \delta ({\vec x}, {\vec y}).
\label{lpb}
\end{equation}
Here the subscript `$L$' denotes the fact that the Poisson bracket 
is for linearised theory.

Note that the flat spatial metric corresponding to the background triad is 
just the Kronecker delta, $\delta_{ab}$. In what follows spatial indices are 
 lowered and raised with this flat metric and its inverse. Internal 
indices are, of course,  
lowered and raised by the $SU(2)$ Cartan-Killing metric. We  also use the
background triad to freely interchange internal and spatial indices.
The flat derivative operator which annihilates the background triad is 
denoted by $\partial_a$. 

The linearised constraints are obtained from 
(\ref{gauss}), (\ref{vector})and (\ref{scalar}) by keeping terms 
at most linear in the
fluctuations and are denoted by $G^L_i, V^L_a,$ and $C^L$  with 
\begin{eqnarray}
G^L_i &=& \partial_a e^a_i + \epsilon_i^{\;ja}A_{aj} , 
\label{lgauss}\\
V^L_a &=& f_{ab}^a,
\label{lvector} \\
C^L &=& \epsilon^{abc} f_{abc}. 
\label{lscalar}
\end{eqnarray}
Here $f_{ab}^i = \partial_a A_b^i-\partial_b A_a^i$ is the linearised
curvature.

The transformations generated by $G^L(\Lambda): =\int d^3x\Lambda^i G^L_i$ are 
\begin{equation}
\delta A_a^i = \{ A_a^i, G^L(\Lambda)\}
 = -\partial_a ({\gamma_0 \Lambda^i\over 2})
\label{lgaussa}
\end{equation}
and 
\begin{equation}
\delta e_i^a = \{ e_i^a, G^L(\Lambda)\} = -\epsilon_i^{\;ak}
                     ({\gamma_0 \Lambda_k\over 2}).
\label{lgausse}
\end{equation}
From (\ref{lgaussa}), $A_a^i$ for each `$i$' transforms as a $U(1)$ connection.
Thus, the configuration space of linearised gravity in this formulation is 
coordinatized by a triplet of $U(1)$ connections $A_a^1,A_a^2,A_a^3$.

In order to construct the loop representation in the next section, we define
the following set of $G^L_i$- invariant  functions on phase space:
\begin{eqnarray}
h^{ab} & = & e^{ab} + e^{ba} ,
\label{h} \\
H^k_{\alpha}&= & \exp i\oint_\alpha A_a^kdx^a .
\label{t}
\end{eqnarray}
Here $\alpha$ is a piecewise analytic, oriented 
loop in $R^3$ and $H^k_{\alpha}$ 
is  the $U(1)$ holonomy of $A_a^k$ around the loop $\alpha$.

It is also useful to define the magnetic field of the connection by 
\begin{equation}
B^a_k =  {1\over 2}\epsilon^{abc} f_{bck}.
\label{B}
\end{equation}
In terms of the magnetic field the vector and scalar constraints are
\begin{eqnarray}
V^L_a &=& \epsilon_{cab}B^{ca}
\label{lvectorb} \\
C^L &=& B^c_{\;c}. 
\label{lscalarb}
\end{eqnarray}
Thus the vanishing of the vector and scalar constraints imply that the 
magnetic field is symmetric and tracefree.

\section*{3. The `loop' representation of quantum linearised gravity}
 
We construct a loop representation based on the $U(1)^3$ holonomies of section
2. The representation at the kinematic ($G^L_i$ invariant) level
is just the tensor product of 3 copies of the $U(1)$ representation worked out in detail in \cite{medual}. We use the notation of, and assume familiarity with
that work.

After presenting the kinematic Hilbert space in section 3a, we turn  our
attention to the linearised vector and scalar constraints in section 3b. Since
the constraints are algebraic statements about the magnetic field, we express  the classical magnetic field via a limit of the holonomy of a shrinking loop
in the usual way. The corresponding quantum operator is not defined on the 
kinematic Hilbert space because the diffeomorphism invariance of the Hilbert 
space measure precludes the existence of the required limit. We show how to 
define the magnetic field operator based on the  dual action of the holonomy operator on a suitable space of 
distributions. We use this defintion to find the kernel of the linearised
vector and scalar constraints.

\subsection*{3a. The kinematic Hilbert space representation}

The kinematic Hilbert space, ${\cal H}_{kin}$,
  inherits its measure from the Haar measure 
on $U(1)$ (it is just the triple product of the 
Ashtekar-Lewandowski measure for $U(1)$ 
connections \cite{AL}). A spanning orthonormal basis is given by the triple
tensor product of the $U(1)$ flux network basis of \cite{medual}.
\footnote{In \cite{medual} this was called the charge network basis; we use
the term flux network to agree with the more recent work \cite{ALfock}.}
Each basis state is labelled by a triplet of closed, oriented,
piecwise analytic graphs as well as  3 sets of 
integers (these are representation labels for $U(1)$),
 one for each graph of the triplet. 
Each set of integers labels the edges of its corresponding graph in such 
a way that at every vertex the sum of labels of outgoing edges equals the sum
of labels of incoming edges. 

We denote the flux network labelled by the graphs $\alpha_i$ and the sets
of integers ${q}_i$, $i=1..3$ as
\begin{equation}
{| \alphaa, \bf{\{q\}} >}
 = |\alpha_1\{q_1\}>|\alpha_2\{q_2\}>|\alpha_3\{q_3\}>.
\label{alphaa}
\end{equation}

As shown in \cite{medual}, the $U(1)$ holonomy of any piecewise analytic loop 
$\beta$ is equally well  associated 
with a $U(1)$ flux network label $\alpha ,\{q\}$ such that 
\begin{equation}
X^a_{\alpha ,\{q\}} ({\vec x}) = X^a_{\beta}({\vec x}).
\end{equation}

Here 
\begin{equation}
X^a_{\beta}({\vec x}):=
\oint_{\beta} ds \delta^3({\vec {\beta}} (s), {\vec x}){\dot{\beta}}^a,
\label{eq:x}                                
\end{equation}
and 
\begin{equation}
X^a_{\alpha ,\{ q\}}({\vec x}) :=  
\sum_{I=1}^N 
q_I\int_{e_I} ds_I \delta^3({\vec e}_I (s_I), {\vec x}){\dot{e_I}}^a .
\label{eq:xe}                                
\end{equation}
where $e_I$ is the $I$th edge of $\alpha$ and is labelled by the integer
$q_I$.

The gauge invariant operators ${\hat H}^i$ and ${\hat h}^{ab}$ are represented
on the kinematic Hilbert space as follows. We first describe the action of 
${\hat H}^1$. This operator acts only on the first ket on the right hand side
of (\ref{alphaa}) exactly as in the case of $U(1)$ theory \cite{medual}. 
Recall, from 
\cite{medual} that, there, the $U(1)$ operator
${\hat H}_{\eta , \{p\}}$ maps
$|\alpha , \{ q\}>$ to a new flux network state based on the 
graph $\alpha \cup \eta$ consisting of the union of the 
sets of  edges belonging
to $\alpha$ and  $\eta$.  
\footnote{It is assumed that edges of $\eta , \alpha$ overlap only if they 
are identical and that intersections of $\eta , \alpha$ occur only at 
vertices of $\eta , \alpha$ . This entails no loss of 
generality, since we can always find graphs which are 
holonomically equivalent to $\eta , \alpha$ and for which the 
assumption holds.}
The edges
of $\alpha \cup \eta$ are oriented and labelled with integers as follows.
Edges which are not shared by $\eta$ and $\alpha$ retain their orientations
and labels. Any shared edge labelled by the integer $q$ in $\alpha$
retains its orientation from $\alpha$ and has the label $q+p$ if it has the 
same orientation in $\eta$ and the label $q-p$ if it has opposite orientation 
in $\eta$. The new state is denoted (with a minor change of notation with
respect to \cite{medual}) by  
$|\alpha , \{q\}\cup \eta ,\{ p\}>$.
This implies that in the $U(1)^3$ case we have,
\begin{equation}
{\hat H}^1_{\eta, \{p\}} {| \alphaa, \bf{\{q\}} >} 
= |\alpha_1,\{q_1\}\cup \eta, \{ p\}>|\alpha_2, \{q_2\}>|\alpha_3, \{q_3\}>.
\end{equation}
Similarly ${\hat H}^2,{\hat H}^3$ act by the union operation on the 
labels $\alpha_2\{q_2\}$ and $\alpha_3\{q_3\}$. Using the notation of
\cite{lingrav}, we write
\begin{equation}
{\hat H}^k_{\eta, \{p\}} {| \alphaa , \bf{\{q\}} >} 
={|\alphaa,  \bf{\{q\}}}\cup_k \eta, \{ p\}> .
\label{lHhat}
\end{equation}
As in \cite{medual} we shall use the labelling of holonomies by their
associated flux network labels (i.e. $H^k_{\alpha \{q\}})$ interchangeably 
with their labelling by loops (i.e. $H^k_{\beta}$). Thus if there is no 
integer label in the subscript to $H$, the label is to be understood as a loop
else as an associated flux network label. 

Also, note that if $\beta$ is a 
loop with a single edge, then the associated flux network label comprises
of the graph $\beta$ with its single edge labelled by the integer $1$. For this
special case we write 
\begin{equation}
H^k_{\beta} = H^k_{\beta, \{1\}}.
\end{equation}

${\hat h}^{ab}$ is represented as
\begin{equation}
{\hat h}^{ab}({\vec x})  {| \alphaa , \bf{\{q\}} >}
= \gamma_0 X^{(ab)}_{{\alphaa} , \bf{\{q\}}}({\vec x})
\label{lhhat}
\end{equation}
where we have defined 
\begin{equation}
 X^{ab}_{\alphaa , \bf{\{q\}}}({\vec x}) = 
\sum_{i=1}^{3}X^{a}_{\alpha_i, \{q_i\}}({\vec x})\delta^{b}_i
\label{defxab}
\end{equation}

It can be verified that (\ref{lHhat}) and (\ref{lhhat})
provide 
a  ${}^*$ 
representation (on the kinematic Hilbert space) 
of the Poisson bracket algebra of the $G^L_i$ invariant 
functions $H^k_{\alpha}$ and $h_{ab}({\vec x})$. Therefore the linearised 
Gauss law constraint is already taken care of and we need  to analyse only   
the remaining  (quantum) vector and scalar constraints.

\subsection*{3b. The Magnetic field operator and physical states}
The magnetic field is extracted from the holonomies of small loops through
\begin{equation}
B^{ck}({\vec x}) = \lim_{\delta\rightarrow 0}{i\over\pi \delta^2}
                             (H^k_{(\gamma^c_{{\vec x}, \delta})^{-1}} -1)
\label{defB}
\end{equation}
where $*$ denotes complex conjugation and $\gamma^c_{{\vec x}, \delta}$
is a   circular loop of radius $\delta$ centered at ${\vec x}$
traversing anticlockwise about and with its plane normal to, the `$c$' axis. 
$(\gamma^c_{{\vec x}, \delta})^{-1}$ denotes the same loop running clockwise.
The corresponding operator 
\begin{equation}
{\hat B}^{ck}({\vec x}) = \lim_{\delta\rightarrow 0}{i\over\pi \delta^2}
                     ({\hat H}^k_{(\gamma^c_{{\vec x}, \delta})^{-1}} -1)
\label{defhatB}
\end{equation}
is not well defined on the finite span of flux network states. The reason is 
that, due to the diffeomorphism invariance of the $U(1)^3$ Ashtekar-Lewandowski
measure, flux network states associated with the triplet of graphs
$\alphaa \cup_k (\gamma^c_{{\vec x}, \delta})^{-1}$  (here we use the notation of \cite{lingrav}) for different values of $\delta$ are orthogonal.

Instead we attempt to define the operator ${\hat B}^{ck}$ by its dual action 
on the space of algebraic duals to the finite span of flux network states.
Recall that the dual 
(anti-)representation of an operator ${\hat A}$ is given by \cite{medual}
\begin{equation}
{\hat A} \Phi(|\psi >) = \Phi ({\hat A}^{\dagger} |\psi >)
\label{dualrep}
\end{equation}
where $\Phi$ is an element of the algebraic dual and $|\psi >$ is a finite linear 
combination of flux network states. Every element of the algebraic dual can
be formally written as an infinite sum over all flux network states i.e.
\begin{equation}
\Phi := \sum_{\alphaa, \bf{\{q\}}} c_{\alphaa, \bf{\{q\}}}
                     <\alphaa, \bf{\{q\}}|
\label{Phic}
\end{equation}
with 
$ c_{\alphaa, \bf{\{q\}}} = \Phi (|\alphaa, \bf{\{q\}}>)$.
It follows that 
\begin{eqnarray}
{\hat B}^{ck}({\vec x}) \Phi &= & \lim_{\delta\rightarrow 0}
             \sum_{\alphaa, \bf{\{q\}}} c_{\alphaa, \bf{\{q\}}}
                     {<{\alphaa}, \bf{\{q\}}|}
    {({\hat H}^{\dagger}_{(\gamma^c_{{\vec x}, \delta})^{-1}} -1)
                 \over i\pi \delta^2} 
\nonumber\\
&= & \sum_{\alphaa, \bf{\{q\}}}\lim_{\delta\rightarrow 0}
       {c_{{\alphaa, \bf{\{q\}}}\cup_k \gamma^c_{{\vec x}, \delta},\{1\}}
            -c_{\alphaa, \bf{\{q\}}}\over {i\pi \delta^2}} 
{<\alphaa, \bf{\{q\}}|}  .
\label{3b0}
\end{eqnarray}
We shall say that ${\hat B}^{ck}({\vec x})$ is well defined iff 
\begin{equation}
{\delta c_{\alphaa, \bf{\{q\}}}\over \delta \gamma^{ck}({\vec x})}:=
\lim_{\delta\rightarrow 0}
       {c_{{\alphaa, \bf{\{q\}}}\cup_k \gamma^c_{{\vec x}, \delta},\{1\}}
            -c_{\alphaa, \bf{\{q\}}}\over {\pi \delta^2}}
\end{equation}
 is well defined. 

We further require that $c_{\alphaa, \bf{\{q\}}}$ is a 
functional of $X^a_{\alpha_i,\{q_i\}}, i=1..3$. This requirement combined with 
the requirement that ${\hat B}^{ck}$ 
be well defined singles out a vector space,
$\Phi^{*L}_{kin}$, of 
`well behaved' distributions.  To summarise: the magnetic field operator is 
defineable, via the dual action of holonomy operators, on the space
$\Phi^{*L}_{kin}$.

The Fourier transform of $X^b_{\alpha_i,\{q_i\}}({\vec x})$  is 
\begin{equation}
X^b_{\alpha_i,\{q_i\}}({\vec p})={1\over (2\pi)^{3\over 2}}\int d^3x
X^b_{\alpha_i,\{q_i\}}({\vec x}) e^{-i{\vec p}\cdot {\vec x}}.
\end{equation}
Define 
\begin{equation}
{\delta X^b_{\alpha_j,\{q_j\}}({\vec p})\over \delta \gamma^{ci}({\vec x})}
:=\delta_{ij}\lim_{\delta\rightarrow 0}
{X^b_{\alpha_j, \{q_j\}\cup\gamma^c_{{\vec x}, \delta},\{1\}}
            -X^b_{\alpha_j, \{q_j\}}\over 
                      {\pi \delta^2}}.
\end{equation}
From  (\ref{eq:xe}) it follows that 
\begin{equation}
{\delta X^b_{\alpha_j,\{q_j\}}({\vec p})\over \delta \gamma^{ci}({\vec x})}
= {-i\over (2\pi)^{3\over 2}} \delta_{ij} p_m \epsilon^{cmb}
e^{-i{\vec p}\cdot {\vec x}}.
\label{3b1}
\end{equation}
Note that 
\begin{equation}
{\delta c_{\alphaa, \bf{\{q\}}}\over \delta \gamma^{ck}({\vec x})}=
\int d^3p {\delta c_{\alphaa, \bf{\{q\}}}\;\;\;\over 
\delta X^b_{\alpha_j,\{q_j\}}({\vec p})}
{\delta X^b_{\alpha_j,\{q_j\}}({\vec p})
\over \delta \gamma^{ci}({\vec x})\;\;\;}.
\label{3b2}
\end{equation}
Using (\ref{3b1}) and (\ref{3b2}) in (\ref{3b0}) and taking the Fourier 
transform of ${\hat B}^{ck}({\vec x})$, we obtain
\begin{equation}
{\hat B}^{ck}({\vec p}) =
\sum_{\alphaa, \bf{\{q\}}}
{\delta c_{\alphaa, \bf{\{q\}}}\over 
\delta X^b_{\alpha_j,\{q_j\}}(-{\vec p})}<\alphaa, \bf{\{q\}}|  .
\label{3b3}
\end{equation}

It is straightforward to show that the constraints in the form 
(\ref{lvectorb}) and (\ref{lscalarb})imply that 
$c_{\alphaa, \bf{\{q\}}}$ depends only on the symmetric, transverse, traceless
(STT)  part of $X^{bc}_{\alphaa, \bf{\{q\}}}$ (the latter is defined in 
(\ref{defxab})). In the standard helicity basis of transverse vectors $m_a,
{\bar m}_a$ (\cite{lingrav}) the STT part of  $X^{bc}_{\alphaa, \bf{\{q\}}}$
can be written as
\begin{equation}
 X^{ab(STT)}_{\alphaa, \bf{\{q\}}}({\vec k})
=X^{+}_{\alphaa, \bf{\{q\}}}({\vec k})m^am^b
+X^{-}_{\alphaa, \bf{\{q\}}}({\vec k}){\bar m}^a{\bar m}^b .
\label{xstt}
\end{equation}
Denote the space of physical states by $\Phi^{*L}_{phys}$. Then 
we have shown that $\Phi \in \Phi^{*L}_{phys}$ iff the coefficients
$c_{\alphaa, \bf{\{q\}}}$ in (\ref{Phic}) are functionals only of 
$X^{+}_{\alphaa, \bf{\{q\}}}({\vec k})$ and 
$X^{-}_{\alphaa, \bf{\{q\}}}({\vec k})$.

\section*{4. The relation between gravitons and states in $\Phi^{*L}_{phys}$}

The abelian Poisson bracket algebra of holonomies plays a crucial role in the 
construction of ${\cal H}_{kin}$ \cite{mefock,medual}. As mentioned in section 1, holonomy operators are not well defined on the standard graviton Fock space.
 However, suitably defined ``smeared holonomies'' {\em are} well defined 
operators on Fock space \cite{mefock,lingrav}. It was noticed in 
\cite{mefock,medual} that for the $U(1)$ case, the algebra of smeared 
holonomies is isomorphic to the holonomy algebra. This isomorphism was
used to construct a representation, indistinguishable 
\footnote{Whether (in the $U(1)$ case) indistinguishable even in principle 
or only practically indistinguishable at scales large compared to the smearing
scale, is discussed in \cite{medual}.}
from the Fock representation, in which holonomies are well defined
operators. Since holonomy  operators are defined in the $U(1)$ loop
representation (called the `qef' representation in \cite{medual})
as well as the new `$r$- Fock' representation in \cite{medual}  (here $r$ is 
a length scale used to define the smearing), it was possible to relate the 
$r$-Fock representation to the loop representation in \cite{medual}. The 
considerations of \cite{medual} can be extended to the case of linearised 
gravity in an obvious and straightforward manner and we shall only
present the main results of such an extension in this section.

In section 4a, we briefly review the standard graviton Fock space 
representation based on  linearised ADM variables. In section 4b, we use the 
Poincare invariance condition \cite{medual} to identify the element of 
$\Phi^{*L}_{phys}$ which corresponds to the $r$- Fock vacuum. We expect that 
 this identification can then be used to relate a suitable 
subspace of 
$\Phi^{*L}_{phys}$ to (a dense subspace of) the $r$- Fock space, modulo 
a couple of open technical issues which we discuss in section 4c.

\subsection*{4a. Review of the standard Fock space representation
of linearised gravity}

The standard Fock representation is obtained by quantizing the true degrees
of freedom in the ADM description. In the ADM description the phase space variables are the linearised metric , $\alpha_{ab}$ and the linearised
ADM momentum, $P^{ab}$ with 
\begin{equation}
\{\alpha_{ab}({\vec x}), P^{cd}({\vec y}) \} = 
\delta_{a}^{(c}\delta_{b}^{d)} \delta ({\vec x},{\vec y}).
\label{4.1}
\end{equation}
The true  degrees of freedom are parametrised by the transverse, traceless 
part of $\alpha_{ab}$ and $P^{cd}$ and are denoted by
$\alpha_{ab}^{TT}$ and $P^{cdTT}$. The true Hamiltonian is 
\begin{equation}
H_L =\int d^3x 
({\partial_m \alpha_{cd}^{TT}\over 2}{\partial^m \alpha^{cdTT}\over 2}
         \;\;   + \;\; P^{cdTT}P_{cd}^{TT} )
\end{equation}
so that 
\begin{equation}
{\dot \alpha}_{cd}^{TT} = 2 P_{cd}^{TT} , \;\;\;\;
{\dot P}_{cd}^{TT}= {\partial ^m \partial_m \alpha_{cd}^{TT}\over 2} .
\label{4.2}
\end{equation}
These evolution equations together imply that 
\begin{equation}
\Box \alpha^{TT}_{cd} =0
\end{equation}
which in turn implies that $\alpha^{TT}_{cd}$ has the following 
plane wave expansion
\begin{eqnarray}
\alpha_{cd}^{TT} ({\vec x}, t) 
& =& {1\over (2\pi)^{3\over 2}}\int {d^3k \over \sqrt{k}}
   ( a_{(+)}({\vec k})e^{i({\vec k}\cdot{\vec x}-kt)}m_cm_d
    + a^*_{(+)}({\vec k})e^{-i({\vec k}\cdot{\vec x}-kt)}{\bar m}_c{\bar m}_d
 \nonumber \\
& & \;\;\;\;\;\;\;\;
 +\; a_{(-)}({\vec k})e^{i({\vec k}\cdot{\vec x}-kt)}{\bar m}_c{\bar m}_d      
    +  a^*_{(-)}({\vec k})e^{-i({\vec k}\cdot{\vec x}-kt)}m_cm_d ).
\label{4.2a}
\end{eqnarray}
Here $k= |{\vec k}|$ and $t$ is the background Minkowskian
time.
From (\ref{4.1}) and (\ref{4.2}) the only non-vanishing Poisson brackets 
between the mode coefficients are
\begin{equation}
\{a_{(\pm)}({\vec k}),a^*_{(\pm)}({\vec l}) \} = -i \delta ({\vec k},{\vec l}).
\label{4.3}
\end{equation}
In quantum theory, ${\hat a}_{(+)}({\vec k})$ and  ${\hat a}_{(-)}({\vec k})$  are represented as
annihilation operators for positive and negative helicity gravitons of  
wave number ${\vec k}$ and ${\hat a}^{\dagger}_{(+)}({\vec k})$ and  
${\hat a}^{\dagger}_{(-)}({\vec k})$  are the corresponding creation 
operators.

\subsection*{4b. $r$-Fock states as elements of $\Phi^{*L}_{phys}$}
It is straightforward to show that the reduced phase space in the 
connection based description of section 2 is naturally coordinatized by 
the symmetric, transverse, traceless part of $A_{ab}$
and the transverse, tracelss part of $h^{ab}$ (recall that $h^{ab}$ is 
symmetric). From (\ref{qab}),(\ref{kab}) and (\ref{ltriad}) it follows that 
\begin{equation}
h^{abTT} = - \alpha^{abTT}
\end{equation}
and that 
\begin{equation}
A_{af}^{STT} = \epsilon_{f}^{\;cd} {\partial_c \alpha^{TT}_{ad}\over 2}
+ \gamma_0 P^{TT}_{af} .
\end{equation}
Using (\ref{4.2a}) to express the Fourier transform of 
 ${\hat A}_{af}^{STT}({\vec x})$ on Fock space in terms of
creation and annihilation operators, we get 
\begin{eqnarray}
{\hat A}_{ab}^{STT}({\vec k})
&=& {\sqrt{k}m_am_b\over 2}({\hat a}_{(+)}({\vec k})[1-i\gamma_0] 
                      +{\hat a}^{\dagger}_{(+)}(-{\vec k})
                         [1+i\gamma_0] \nonumber \\  
&+& {\sqrt{k}{\bar m}_a{\bar m}_b\over 2}
    ({\hat a}_{(-)}({\vec k})[-1-i\gamma_0] 
                      +{\hat a}^{\dagger}_{(-)}(-{\vec k})
                         [-1+i\gamma_0] ).
\label{atoadm}
\end{eqnarray}

We define the smeared holonomy (also called the $r$-holonomy) labelled by 
$\alphaa, \bf{\{q\}}$ as
\begin{equation}
H^{STT}_{\alphaa, \bf{\{q\}}(r)}:= \exp 
i\int d^3 k X^{ab}_{\alphaa, \bf{\{q\}}(r)}(-{\vec k})A_{ab}^{STT}({\vec k})
\label{4.4}
\end{equation}
where 
\begin{equation}
X^{ab}_{\alphaa, \bf{\{q\}}(r)}({\vec k})
= e^{-k^2r^2\over 2}X^{ab}_{\alphaa, \bf{\{q\}}}({\vec k}).
\end{equation}
Poincare invariance is fed into the construction of the Fock space representation through the specific choice of complex structure (i.e. the positive- negative frequency decomposition (\ref{4.2a})). This choice is equivalent to the
requirement that the Fock vacuum be a zero eigenstate of  the
annhilation operators. This requirement can, in turn, be encoded in terms of 
the smeared holonomy operators as
\begin{eqnarray}
\exp \big(
{i{\gamma_0\over 4}\int d^3x X^{ab}_{\alphaa, \bf{\{q\}}(r)}({\vec x})
                G_{ab}^{\alphaa, \bf{\{q\}}(r)}({\vec x})}
\big)&
\exp \big(
{{i\over 2}\int d^3x G_{ab}^{\alphaa, \bf{\{q\}}(r)}({\vec x})
{\hat h}^{ab}({\vec x})}
\big) |0> \nonumber \\
&  = {\hat H}^{STT}_{\alphaa, \bf{\{-q\}}(r)} |0>               
\label{fockvac}
\end{eqnarray}
where $G_{ab}^{\alphaa, \bf{\{q\}}(r)}({\vec x})$ is defined through its 
Fourier transform,
\begin{equation}
G_{ab}^{\alphaa, \bf{\{q\}}(r)}({\vec k}) 
= k X^+_{\alphaa, \bf{\{q\}}(r)}({\vec k}) (1+i\gamma_0)m_am_b
- k X^-_{\alphaa, \bf{\{q\}}(r)}({\vec k}) (1-i\gamma_0){\bar m}_a{\bar m}_b
\end{equation}
The image of this condition in the $r$-Fock representation is 
\begin{eqnarray}
\exp\big(
{i{\gamma_0\over 4}\int d^3x X^{ab}_{\alphaa, \bf{\{q\}}(r)}({\vec x})
                G_{ab}^{\alphaa, \bf{\{q\}}(r)}({\vec x})}
\big)
 & \exp\big({{i\over 2}\int d^3x G_{ab}^{\alphaa, \bf{\{q\}}(r)}({\vec x})
{\hat h}^{ab}_{r}({\vec x})} |0_r>
\big)\nonumber \\
&=  {\hat H}^{STT}_{\alphaa, \bf{\{-q\}}} |0_r>              
\label{rfockvac}
\end{eqnarray}
where 
\begin{equation}
h^{ab}_{r}({\vec k})=e^{-k^2r^2\over 2}h^{ab}({\vec k})
\end{equation}
and 
\begin{equation}
  H^{STT}_{\alphaa, \bf{\{q\}}}
=\exp i\int d^3x X^{ab}_{\alphaa, \bf{\{q\}}(r)}({\vec x})
A_{ab}^{STT}({\vec x}).
\label{hsttdef}
\end{equation}
The $r$-Fock vacuum bra, $<0_r|$, can be identified with the element
$\Phi_0 \in \Phi^{*L}_{phys}$  via the following equation in the dual
representation (see (\ref{dualrep})). Let $|\psi >$ be a finite
linear combination of flux network states. Then 
\begin{eqnarray}
\Phi_0\bigg(
\exp\big(
{-i{\gamma_0\over 4}\int d^3x X^{ab}_{\alphaa, \bf{\{q\}}(r)}({\vec x})
                (G_{ab}^{\alphaa, \bf{\{q\}}(r)}({\vec x})})^*
\big)
&\exp\big(
{-{i\over 2}\int d^3x (G_{ab}^{\alphaa, \bf{\{q\}}(r)}({\vec x}))^*
{\hat h}^{ab}_{r}({\vec x})} |\psi>\bigg) \nonumber \\
&=  \Phi_0({\hat H}^{\dagger}_{\alphaa, \bf{\{-q\}}} |\psi >)   .           
\label{4.5}
\end{eqnarray}
Here ${\hat H}^{\dagger}_{\alphaa, \bf{\{-q\}}}$ is defined through
\begin{eqnarray}
H_{\alphaa, \bf{\{q\}}} &= & \exp i\int d^3x 
X^{ab}_{\alphaa, \bf{\{q\}}}({\vec x})A_{ab}({\vec x})\nonumber\\
&=& \prod_{k=1}^3\exp i\int d^3x 
X^{a}_{\alpha_k, \{q_k\}}({\vec x})A_{a}^k({\vec x}).
\end{eqnarray}
Note that in (\ref{4.5}) we have effectively replaced $A_{ab}^{STT}$ in 
(\ref{4.4})by $A_a^i$. This is correct because the operator 
${\hat H}_{\alphaa, \bf{\{-q\}}}$ is defined on {\em physical states}.
Since such states are in the kernel of the constraints and since
$H_{\alphaa, \bf{\{-q\}}}$ is a Dirac observable, the $STT$ condition
is automatically
enforced on $\Phi^{*L}_{phys}$. 

As in (\ref{Phic}) we set
\begin{equation}
\Phi_0 := \sum_{\alphaa, \bf{\{q\}}} c_{0\alphaa, \bf{\{q\}}}
                     <\alphaa, \bf{\{q\}}|
\label{Phic0}
\end{equation}
and solve for the coefficients $c_{0\alphaa, \bf{\{q\}}}$ The unique (upto 
an overall multiplicative constant) solution is 
\begin{equation}
c_{0\alphaa, \bf{\{q\}}}
= \exp\big({-i{\gamma_0\over 4}\int d^3x 
G_{ab}^{\alphaa, \bf{\{q\}}(r)}({\vec x})})^*
X^{ab}_{\alphaa, \bf{\{q\}}(r)}({\vec x})
\big).
\end{equation}

\subsection*{4c. Open technical issues}
In \cite{mefock}, it was shown that the set of states obtained by the 
action of the holonomy operators on the $r$-Fock vacuum is dense in the
$r$-Fock space. Denote this set by $D$. A corresponding set of distributions,
${\cal D}^*$, in the (dual) loop representation was obtained by the dual
action of the holonomy operators on $\Phi_0$.
The inner product between two elements of ${\cal D}^*$ was defined to be 
equal to the $r$-Fock inner product between the two corresponding elements of 
$D$ (see (45) of \cite{medual}). This procedure is consistent provided the 
set of distributions in ${\cal D}^*$ corresponding to 
any (finite) linearly independent set of vectors in $D$, is linearly 
independent in ${\cal D}^*$. A cursory glance at this provisio indicates 
that its validity is very plausible  but a proof, as yet, does not exist.
\footnote{We did not realise the necessity of proving this in \cite{medual}.}

In the case of linearised gravity, it should be straightforward to generalise
the results of \cite{mefock} to show that the set of states obtained by 
the action of the operators 
${\hat H}^{STT}_{\alphaa, \bf{\{q\}}}$ on $|0_r>$ generates a dense subspace,
$D_{r-Fock}$, of the $r$-graviton Fock space. The corresponding set, 
$\Phi^{*L}_{r-Fock}$ can be identified by the dual action of 
${\hat H}_{\alphaa, \bf{\{q\}}}$ on $\Phi_0$ and the inner product on
$\Phi^{*L}_{r-Fock}$ can be induced from that on $D_{r-Fock}$ by
\begin{equation}
({\hat H}_{\alphaa, \bf{\{q\}}}\Phi_0, {\hat H}_{\betaa, \bf{\{p\}}}\Phi_0)
= <0_r|{\hat H}^{\dagger STT}_{\betaa, \bf{\{p\}}} 
{\hat H}^{STT}_{\alphaa, \bf{\{q\}}}|0_r> 
\label{innerprod}
\end{equation}
Further, $\Phi^{*L}_{r-Fock}$ can be completed to a Hilbert space naturally 
isomorphic to the $r$-Fock space.
Again, the procedure is consistent provided every finite linearly independent
set of vectors in $D_{r-Fock}$ defines a corresponding 
linearly independent set of vectors in $\Phi^{*L}_{r-Fock}$. This remains 
to be shown but seems to be quite plausible. 

We close with some remarks on the incorporation of the 
reality properties of the phase space variables in terms of 
adjointness properties of appropriate quantum operators. 

Modulo the open issues above, note that:\\
\noindent (1)  the operators ${\hat H}_{\alphaa, \bf{\{q\}}}$ and  
\begin{equation}
{\hat M}_{\betaa, \bf{\{p\}}(r)}:=
\exp \big(
{{i\over 2}\int d^3x G_{ab}^{\betaa, \bf{\{p\}}(r)}({\vec x})
{\hat h}_{(r)}^{ab}({\vec x})}
\big) .
\end{equation}
provide an (anti-)
 representation on $\Phi^{*L}_{phys}$
of the corresponding Poisson bracket  algebra. 

\noindent (2)  the action 
on  $\Phi_0$  of the operator 
${\hat M}_{\alphaa, \bf{\{q\}}(r)}$ 
is uniquely determined 
in terms of that of ${\hat H}_{\alphaa, \bf{\{q\}}}$ 
from (\ref{4.5}).
This, in conjunction with (1),
implies that the action of the operators 
${\hat H}_{\alphaa, \bf{\{q\}}}$ and ${\hat M}_{\betaa, \bf{\{p\}}(r)}$
on $\Phi^{*L}_{r-Fock}$
is naturally isomorphic to the action of the corresponding operators on 
$D_{r-Fock}$ in 
the (dual) $r$-Fock representation.

\noindent (3)the $r$-Fock inner product correctly enforces 
the adjointness properties of these operators in the $r$-Fock representation.

From (1)- (3) above, it is 
reasonable to expect that the inner product (\ref{innerprod}) incorporates
the appropriate reality conditions. However, an explicit proof of this is 
still lacking and is expected to be a bit involved for the following reason.
The function $G_{ab}^{\alphaa, \bf{\{q\}}(r)}({\vec x})$ is complex.
As a result, 
${\hat M}_{\alphaa, \bf{\{q\}}(r)}$  is neither unitary nor hermitian and 
consequently it is expected that the algebra of operators 
generated by ${\hat M}_{\alphaa, \bf{\{q\}}(r)}$ and 
${\hat H}_{\alphaa, \bf{\{q\}}}$  is not closed under the adjoint operation,
thus complicating the required proof.

%Another issue to be clarified arises in the context of linearised gravity due
%to the presence of the linearised vector and scalar constraints. Although
%we expect ${\hat H}_{\alphaa, \bf{\{q\}}}$ to generate the dense set,
%$\Phi^{*L}_{Fock}$, the `individual' holonomies, 
%${\hat H}^j_{\alpha , \{q\}}$, also have a well defined action on $\Phi_0$
%(and indeed, since they are Dirac observables, on the whole of 
%$\Phi^{*L}_{phys}$.). 
%We believe that the state
%${\hat H}^j_{\alpha, \{q\}}\Phi_0$ corresponds to the state
%generated by the action, on $|0_r>$, of the operator corresponding to
%the classical function
%$\exp i\int d^3x X^a_{\alpha, \{q\}} A_{aj}^{STT}$ in the
%$r$-Fock representation.
%This should also not be hard to prove but is still an open issue.

\section*{5. The case of complex $\gamma_0$}
Our considerations till now have been based on the real $SU(2)$ formulation 
of gravity. Remarkably, much of our analysis can also be applied to the 
formulation of section 2 with an arbitrary {\em complex} Barbero- Immirzi
parameter, $\gamma_0$, including the case of $\gamma_0 =-i$ which corresponds
to the choice of self dual variables \cite{selfdual,lingrav}.

We adopt the viewpoint that the kinematic $U(1)^3$  based Hilbert space,
${\cal H}_{kin}$, is simply an auxilliary structure whose only role is 
to furnish a (dual) representation of the algebra ({\em not} the * algebra)
generated by $H_{\alphaa, \bf{\{q\}}}$ and $h^{ab}$. This representation
is to be used to find the kernel of the quantum constraints and the physical
inner product is to be chosen in such a way as to enforce the * relations on 
Dirac observables as adjointness relations on the corresponding  operators.

To this end, the analysis of sections 2 and 3 holds with a complex $\gamma_0$.
Note that the dual representation is defined by (\ref{dualrep}) with the
adjoint operation taken {\em with respect to the kinematic Hilbert space 
inner product}. For $\gamma_0$ complex, 
this `kinematic' adjoint operation does {\em not} enforce 
the * relations obtained from  the `reality conditions' \cite{lingrav}.
The reality conditions on the linearised variables are
\begin{equation}
(h^{ab})^* = h^{ab} \;\;\;\;\;\; 
\big({A_{(ab)} -\Gamma_{(ab)}\over \gamma_0}\big)^*
= \big({A_{(ab)}-\Gamma_{(ab)}\over \gamma_0}\big) 
\label{5.1}
\end{equation}
and are to be incorporated in the quantum theory  by the physical
inner product, not necessarily the kinematic one. In fact, with respect 
to the kinematic adjoint operation, ${\hat h}^{ab}$ is not self adjoint.
Instead in contrast to (\ref{lhhat}) we have that
\begin{equation}
{\hat h}^{\dagger ab}({\vec x})  {| \alphaa , \bf{\{q\}} >}
= \gamma_0^* X^{(ab)}_{{\alphaa} , \bf{\{q\}}}({\vec x})
{| \alphaa , \bf{\{q\}} >}
\label{5.2}
\end{equation}

The contents of section 4b upto and including (\ref{hsttdef}) are valid even
for complex $\gamma_0$. In particular, the Poincare invariance of the vacuum 
is still encoded in (\ref{fockvac}). Equation (\ref{rfockvac}) too, is
unchanged but (\ref{4.5}) in the dual representation must be defined through
(\ref{dualrep}). Since $X^{(ab)}_{{\alphaa} , \bf{\{q\}}}({\vec x})$ is 
real and since (with the kinematic adjoint) 
${\hat h}^{\dagger ab}\neq {\hat h}^{ab}$
when $\gamma_0$ is complex, (\ref{4.5}) is replaced by 
\begin{eqnarray}
\Phi_0\bigg(
\exp\big(
{-i{\gamma^*_0\over 4}\int d^3x X^{ab}_{\alphaa, \bf{\{q\}}(r)}({\vec x})
                (G_{ab}^{\alphaa, \bf{\{q\}}(r)}({\vec x})})^*
\big)
&\exp\big(
{-{i\over 2}\int d^3x (G_{ab}^{\alphaa, \bf{\{q\}}(r)}({\vec x}))^*
{\hat h}^{\dagger ab}_{r}({\vec x})} |\psi>\bigg) \nonumber \\
&=  \Phi_0({\hat H}^{\dagger}_{\alphaa, \bf{\{-q\}}} |\psi >)   .           
\label{5.2a}
\end{eqnarray}
This equation admits the unique (upto an overall multiplicative constant)
solution
\begin{equation}
c_{0\alphaa, \bf{\{q\}}}
= \exp\big({-i{\gamma^*_0\over 4}\int d^3x 
(G_{ab}^{\alphaa, \bf{\{q\}}(r)}({\vec x})})^*
X^{ab}_{\alphaa, \bf{\{q\}}(r)}({\vec x})
\big).
\end{equation}
When  $\gamma_0\neq \pm i$ , we again expect 
the steps of section 4c to go through with the inner product on 
$\Phi^{*L}_{Fock}$ specified through
\begin{equation}
({\hat H}_{\alphaa, \bf{\{q\}}}\Phi_0, {\hat H}_{\betaa, \bf{\{p\}}}\Phi_0)
= <0_r|{\hat H}^{\dagger STT}_{\betaa, \bf{\{p\}}} 
{\hat H}^{STT}_{\alphaa, \bf{\{q\}}}|0_r> 
\label{innerprodcomp}
\end{equation}
where ${\hat H}^{\dagger STT}_{\betaa, \bf{\{p\}}}$ is the adjoint with
respect to the $r$- Fock inner product. The latter correctly incorporates the 
reality conditions given by (\ref{5.1}).
In particular, since $\gamma_0$ is complex, 
${\hat H}^{ STT}_{\betaa, \bf{\{p\}}}$ is {\em not} a unitary operator.
Note that
the comments  in section 4c regarding the incorporation of 
reality conditions in terms of adjointness conditions
also apply to the inner product (\ref{innerprodcomp}).

When $\gamma_0 =i$ or $-i$, (\ref{atoadm}) implies that 
${\hat A}_{ab}^{STT}({\vec k})$ lacks either the positive helicity 
creation operator or the negative helicity creation operator. Hence 
${\hat H}^{STT}_{\alphaa, \bf{\{q\}}(r)}$ cannot generate the positive 
helicity (respectively, negative helicity) graviton sector from the 
vacuum.  Instead, operators involving the linearised metric would have to be 
used to generate the Hilbert space from the vacuum. Although we have not 
attempted the relevant analysis, we do expect that the methods of 
\cite{lingrav} can be recast in the language of this paper to successfully 
do so.

\section*{6. Discussion}
In this work we have shown how the $r$-Fock representation for linearised 
gravity can be constructed, starting from the `loop' representation
on the kinematic Hilbert space, ${\cal H}_{kin}$. The role of the 
representation on  ${\cal H}_{kin}$ in this construction is that it
provides the structure to define, with mathematical precision, the 
dual (anti-)representation on an appropriate space of distributions.
In particular, the role of the kinematic Hilbert space inner product
is to define the kinematic adjoint operation which is, in turn, used to define
the dual representation through (\ref{dualrep}).

One of the features of this work is that it highlights the importance of the 
dual representation on the space of distributions. 
Physical states (as opposed to kinematic ones) lie in this space. The condition
(\ref{rfockvac}) which is satisfied by the $r$- Fock vacuum in the $r$-Fock
representation not only makes sense (in the form of (\ref{4.5})), but also 
admits an essentially unique solution, $\Phi_0$, in the dual representation.
Modulo the comments in  sections 4c and 5,
the rest of (a dense subspace of) the $r$- Fock space is then generated 
from $\Phi_0$, once again, via the dual representation of appropriately 
chosen Dirac observables.

Another feature of this work is that  the inner product  on physical 
states, namely the $r$- Fock inner product, is very different from the 
kinematic inner product. Indeed, the physical states are not kinematically 
normalizable. The results of section 5 further de-emphasize the 
{\em physical}  significance of the kinematic inner product and seem 
to strengthen the old viewpoint in the loop quantum gravity approach wherein 
the physical inner product is to be determined by the reality conditions.
Note, however, that the kinematic structures continue to play a key 
{\em mathematical} role in defining the 
dual representation even when $\gamma_0$
is complex. Thus, even though the rigorous mathematical structures of 
\cite{AI,M^2,AL,ALM^2T} are defined only for compact gauge groups, we were 
able to use such structures profitably, even for the self dual description 
of linearised gravity.

We now turn to a brief discussion of the physical indistinguishability 
of the $r$-Fock and the Fock representations.
In the $U(1)$ context we noted in \cite{medual} that there were two possible
viewpoints with regard to this issue. One viewpoint is that only 
{\em algebraic} properties of functions on phase space are measurable.
This viewpoint applied to linearised gravity would imply that there is no
way of asserting whether the pair
$(H^{STT}_{\alphaa, \bf{\{q\}}(r)}, h^{abSTT})$ is  being measured  in the
Fock representation or the pair  
$(H^{STT}_{\alphaa, \bf{\{q\}}}, h^{abSTT}_{(r)})$ is being measured in the
$r$-Fock representation. Thus, with this viewpoint, the physics of the 
$r$-Fock representation is exactly (not approximately) {\em identical} to 
that of the Fock representation.

The other viewpoint is valid in the case that there is some property
other than purely algebraic properties of the pair
$(H^{STT}_{\alphaa, \bf{\{q\}}(r)}, h^{abSTT})$ by virtue of which 
the measuring apparatus measures them  rather than the pair,
$(H^{STT}_{\alphaa, \bf{\{q\}}}, h^{abSTT}_{(r)})$.  In such a case,
the $r$-Fock representation is physically indistinguishable from the 
Fock representation only for finite accuracy measurements at distance scales 
much larger than $r$ \cite{medual}. Linearised gravity is a truncation 
of full general relativity. In the latter, the primary object which is measured
is the full metric. 
The notion of smearing does not extend to an arbitrary metric in any
natural way (note that the smearing we use is heavily dependent on the 
background flat metric). Thus, for a reason external to the narrow
confines of linearised gravity, we expect that the physical apparatus
measures the combination $\delta^{ab} + h^{ab}$ from which $h^{ab}$ can 
be estimated. Hence the object $h^{ab}$ rather than $h^{ab}_{(r)}$ is
preferred and the second viewpoint mentioned above seems to be the valid one.
We have explored consequences  of this viewpoint for violation of 
Poincare invariance at scales smaller than $r$ and will report our results
elsewhere \cite{dispersion}.

As mentioned in the introduction, the deeper question of how (if at all!)
the $U(1)^3$ loop representation arises from loop quantum gravity is as 
yet unsolved. A small preliminary step in this direction would be to 
investigate if the linearised constraints can be solved via an 
`averaging' procedure \cite{raq} similar to that used in loop quantum gravity
\cite{ALM^2T}, rather than by using the magnetic field operator. This would 
bring the $U(1)^3$ approach structurally even
 closer to the loop quantum gravity approach.

This work represents the culmination of our efforts, initiated in 
\cite{mefock} and continued in \cite{medual}, to understand the older results 
of \cite{lingrav} in the mathematically precise 
language currently used in the field. We hope that this work may aid 
current efforts to construct semiclassical states in loop quantum gravity
\cite{ALfock,thomas} and suggest that it may be a profitable venture 
to revisit the older efforts of Iwasaki and Rovelli \cite{junichi} in the 
light of subsequent developments in the field. 

\noindent {\bf Acknowledgements:} I would like to thank
Professor Kleber and
Professor Sackmann  for providing a wonderful research environment  
at the Technische Universit{\" a}t, M{\" u}nchen (Garching),
 where a large part of this work
was completed.

\end{document}